\documentclass[aps,prl,twocolumn,showpacs,preprintnumbers,amsmath,amssymb]{revtex4}
\usepackage[]{graphicx}

\newcommand{\ket}[1]{\mbox{\ensuremath{|#1\rangle}}}

\begin{document}
\begin{abstract}
Since, in general, non-orthogonal states cannot be cloned, any eavesdropping attempt in a Quantum
Communication scheme using non-orthogonal states as carriers of information introduces some errors
in the transmission, leading to the possibility of detecting the spy. Usually, orthogonal states
are not used in Quantum Cryptography schemes since they can be faithfully cloned without altering
the transmitted data. Nevertheless, L. Goldberg and L. Vaidman [\prl 75 (1995) 1239] proposed a
protocol in which, even if the data exchange is realized using two orthogonal states, any attempt
to eavesdrop is detectable by the legal users. In this scheme the orthogonal states are
superpositions of two localized wave packets travelling along separate channels. Here we present an
experiment realizing this scheme.
\end{abstract}
\title{Experimental quantum cryptography scheme based on orthogonal states}
\author{Alessio Avella, Giorgio Brida, Ivo Pietro Degiovanni, Marco Genovese, Marco Gramegna and Paolo Traina}
\affiliation{INRIM, Strada delle cacce 91, Turin, Italy.}
\maketitle






Quantum Key Distribution (QKD) is a method for transmitting a secret key between two partners
(usually named Alice and Bob) by exploiting quantum properties of light. The most important
characteristic of this idea is that the secrecy of the generated key is guaranteed by the very laws
of nature, i.e. by the properties of quantum states \cite{gis}. In the last decade QKD has
abandoned the laboratories becoming a mature technology for commercialization \cite{l};
communications over more thane 100 km having been achieved both in fiber \cite{fib} and open air
\cite{op}.

Various protocols for realising QKD have been suggested \cite{gis},
such as BB84 \cite{BB84}, B92 \cite{B92}, Ekert \cite{ek}. All
of them are based on the use of non-orthogonal states, a condition
that was considered necessary for guaranteeing security, up to a
paper of Goldenberg and Vaidman (GV) \cite{g}. In that work was presented
for the first time a scheme for realising a QKD protocol based on
orthogonal states, whose security was based on two ingredients.
First, the orthogonal states sent by Alice were superposition of two
localised wave packets that were not sent simultaneously to Bob.
Second, the transmission time of the photons was random.

This scheme, beyond its interest for application, has also a large
conceptual interest for understanding the quantum
resources/properties needed for QKD. Nevertheless, up to now, no
experimental realization was still done. Purpose of this letter is
to present its first experimental implementation.

In the theoretical proposal of Ref.\cite{g}, the orthogonal states sent by Alice are the
superpositions of two localized wave-packets. Those  are not sent simultaneously to Bob, but
separated by a fixed delay. In this case there is a direct correspondence between the state
prepared by Alice and the bit received by Bob, for instance,
\begin{eqnarray}
\label{eq:states}
0  &\rightarrow \ket{\Psi_0}&=\frac{1}{\sqrt{2}}(\ket{a}+\ket{b})\nonumber \\
1 &\rightarrow \ket{\Psi_1}&=\frac{1}{\sqrt{2}}(\ket{a}-\ket{b}),\nonumber
\end{eqnarray}
where $\ket{a}$ and $\ket{b}$ are two localized wave--packets and
the states $\ket{\Psi_0}$ and $\ket{\Psi_1}$ are orthogonal. The
states $\ket{\Psi_0}$ and $\ket{\Psi_1}$ are emitted randomly in
time, and the presence of an eventual eavesdropper can be detected
by legitimate users exploiting the information on the detection
times \cite{g}.
The scheme works as follows: Alice sends Bob either $\ket{\Psi_0}$ or $\ket{\Psi_1}$.
The launch on the quantum channel of the wave-packet $\ket{b}$ is delayed for some amount of time
$\tau$ with respect to the launch of wave-packet $\ket{a}$. $\tau$ could be chosen larger than the
traveling time $T$ of photons between Alice's and Bob's locations. As $\ket{b}$ will travel through
the quantum channel only after the wave-packet $\ket{a}$ has already reached Bob's location, both
packets are never simultaneously present in the quantum channels. Furthermore, as pointed out in
Ref. \cite{g}, the requirement of $\tau$ greater than the traveling time $T$ is not strictly
necessary. Indeed the security of the protocol is ensured even if $\tau$ is only greater than the
overall uncertainty in the measurement of the transmission/detection times $t_{s}$ and $t_{r}$
\cite{g}.

In our proof-of-principle experiment this is obtained exploiting a
balanced Mach-Zehnder Interferometer (MZI) with two equal optical delays
$OD_{1}$ and $OD_{2}$. According to Fig. 1, sources of single photon
$S_{0}$ and $S_{1}$ at the two input ports of the beam splitter on
Alice side provide single photons propagating in the transmission
channel in the state $\ket{\Psi_0}$ or $\ket{\Psi_1}$ respectively.
The emission time of the single photon in one of the two state is
random, but it is registered by Alice.

As the packet $\ket{b}$ is stored in $OD_{1}$, wave-packet $\ket{a}$ travels from Alice's to Bob's
site along the upper channel and enters in $OD_{2}$, where it is delayed until also $\ket{b}$
reaches Bob's site. In this way the two packets interfere as they simultaneously  arrive to the
second beam-splitter, thus, the click of detector $D_{i}$ deterministically implies that the single
photon state was in the state $\ket{\Psi_{i}}$, i.e. it was sent by source $S_i$.
Two security tests are performed by Alice and Bob to highlight the possible presence of an
eavesdropper. The first one is a public comparison between the sending times ${t_{s}}$ and the
receiving times ${t_{r}}$ for each photon. If we assume that the traveling time between the two
parties is $T$, only the events detected at time $t_{r}=t_{s}+\tau+T$ are considered
as part of the message,
 while all the
others
highlight the presence of Eve. The second one is the comparison of corresponding portions of the legitimate
users' bit strings to estimate the quantum bit error rate (QBER). We underline that in the ideal
case discrepancies in the transmission/detection times or in the bit strings can only be induced by
an eavesdropper.

Let us mention for the sake of completeness that it was argued by Peres\cite{peres}
that this protocol introduced no novelty with respect to BB84. To this claim GV 
replied that while in other protocols like BB84 the security is guaranteed by nonorthogonality,
in GV protocol it is based on causality, since they proved that a successful eavesdropping
would require superluminal signaling \cite{g}.
Furthermore, while all cryptographic schemes require two steps for sending information
(sending the quantum object and then some classical information), in GV protocol
only the first step is needed for communication, the second step is used only for
assuring security against eavesdropping.
\begin{figure}
   \begin{center}
   \includegraphics[bb=0 0 798 597, width=7cm]{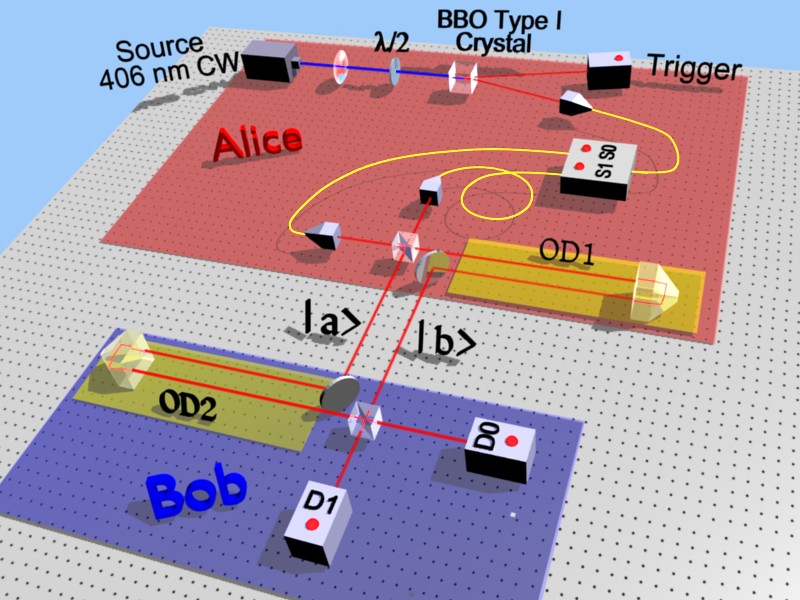}
   \end{center}
\caption{ \label{fig:example2} Experimental set-up
. Two single photon sources ($S_{0}, S_{1}$, realized exploiting an heralded single photon source based on
parametric down-conversion (PDC) obtained by
pumping with a $406$ nm CW laser beam a type I BBO crystal) are injected in the MZI, one at the time.  
Alice's site is composed by the two single photon sources and the first optical delay ($OD_{1}$).
Bob's site is composed by the second optical delay ($OD_{2}$, identical to $OD_{1}$)
and the two single photon detectors ($D_{0}, D_{1}$).}
\end{figure}

Fig.1 shows the setup of the experiment representing the first realization of the GV
protocol. The single photon states are obtained exploiting an heralded single photon source based
on parametric down-conversion (PDC) \cite{mandel}. A CW $100$ mW Coherent Cube diode laser system
at $406$ nm is used to pump a BBO type I crystal. PDC photons pairs at degeneracy ($812$ nm) are
emitted in slightly non-collinear regime (three degrees with respect to the pump direction). The
heralding photons are selected by means of $1$ nm bandwidth interference filters, collected in a
multimode optical fiber and detected by Single Photon Avalanche Photo-diodes (SPAD) detectors. The
heralded single photon, the carrier of the information to be exchanged between the legitimate
parties, is collected in a single mode optical fibre (a $10$ nm interference filter is placed on
the heralding arm only for background suppression). The CW laser operation ensures the generation
of photon pairs at random time, and the detection of one photon of the pair in the heralding arm
provides the temporal information on the emission of the single-photon as requested by the GV
protocol.
With the aim of realizing the proof-of-principle QKD scheme proposed by GV, Alice sends bit
$0$ or $1$ by addressing the encoding photon to the proper input port of the first beam-splitter.
Bob detect the single photons at the output of the interferometer.
The balanced MZI 
contains both the optical delays and the transmission
channel from Alice to Bob. In particular, after the input BS at the Alice side one arm of the
interferometer contains a delay line (realized through a trombone prism), while on the other arm the
delay line (again based on a trombone prism) is located on Bob's side. The positions of the
trombones in the optical delays are adjusted via a closed loop piezo-movement system with
nano-metric resolution. Detection events after the output BS of the interferometer are obtained by
SPAD detectors operating in Geiger mode. The electronics highlighting the presence of coincident
detections is based, as usual, on Time-to-Amplitude-Converter and Multi-Channel-Analyser.
Specifically, in our case the temporal condition for the security of the QKD scheme is satisfied as
the jitter of our detectors (corresponding to the uncertainty in the determination of the of the
transmission/detection times) is about 300 ps, while the length of the delay lines is 60 cm
corresponding to a storage time of $\sim$2 ns.
%
%
The stability of the interferometer has been tested by scanning the position of Alice's trombone
prism with Bob's one kept at a fixed position. Fig. 2 shows the interference fringes of heralded
counts and the visibilities (V) are well above $80 \%$, irrespective of which port of the input
beam splitter is used to inject the single photon in the interferometer.
\begin{figure}
   \begin{center}
   \includegraphics[width=8.2cm]{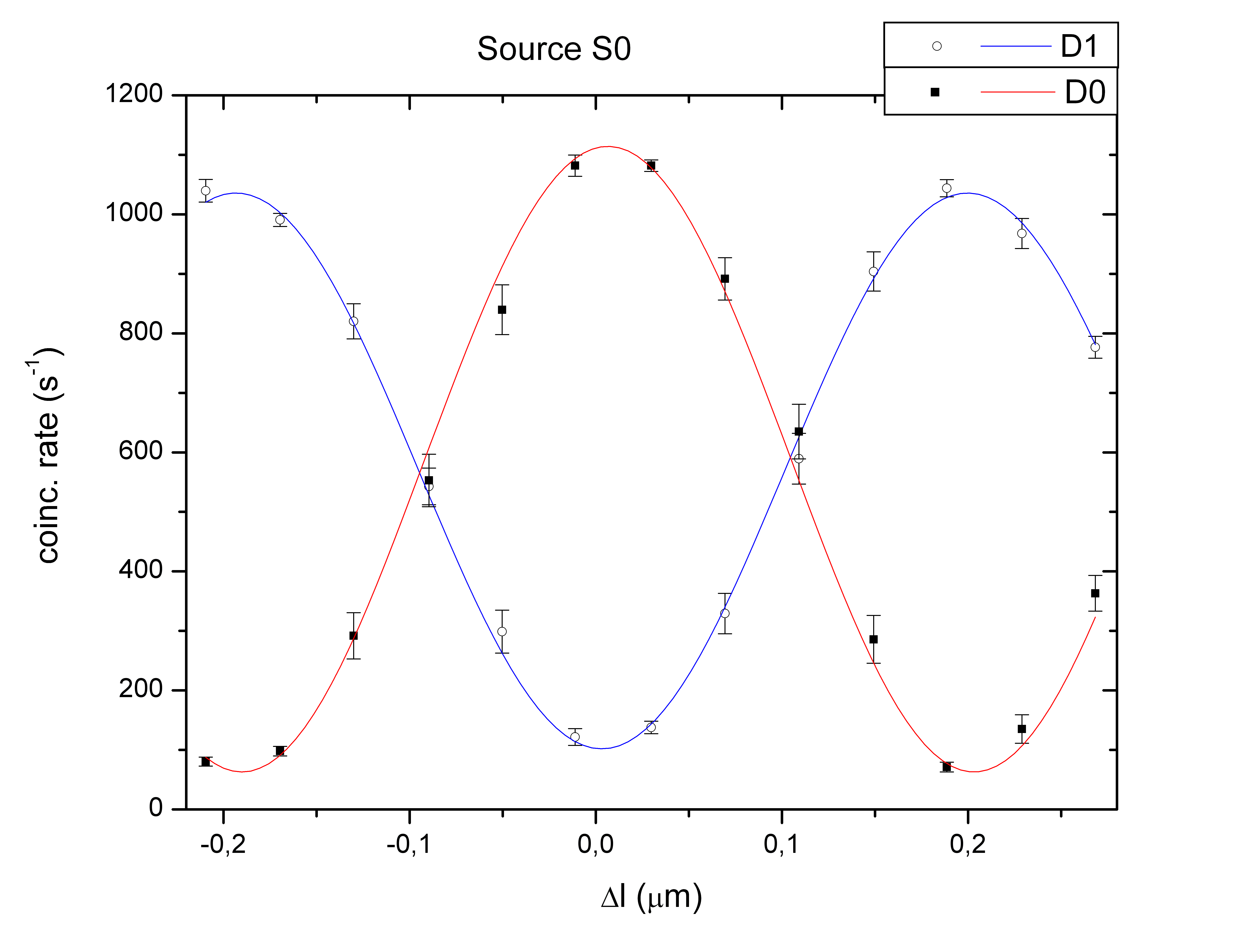}\\ \includegraphics[width=8.2cm]{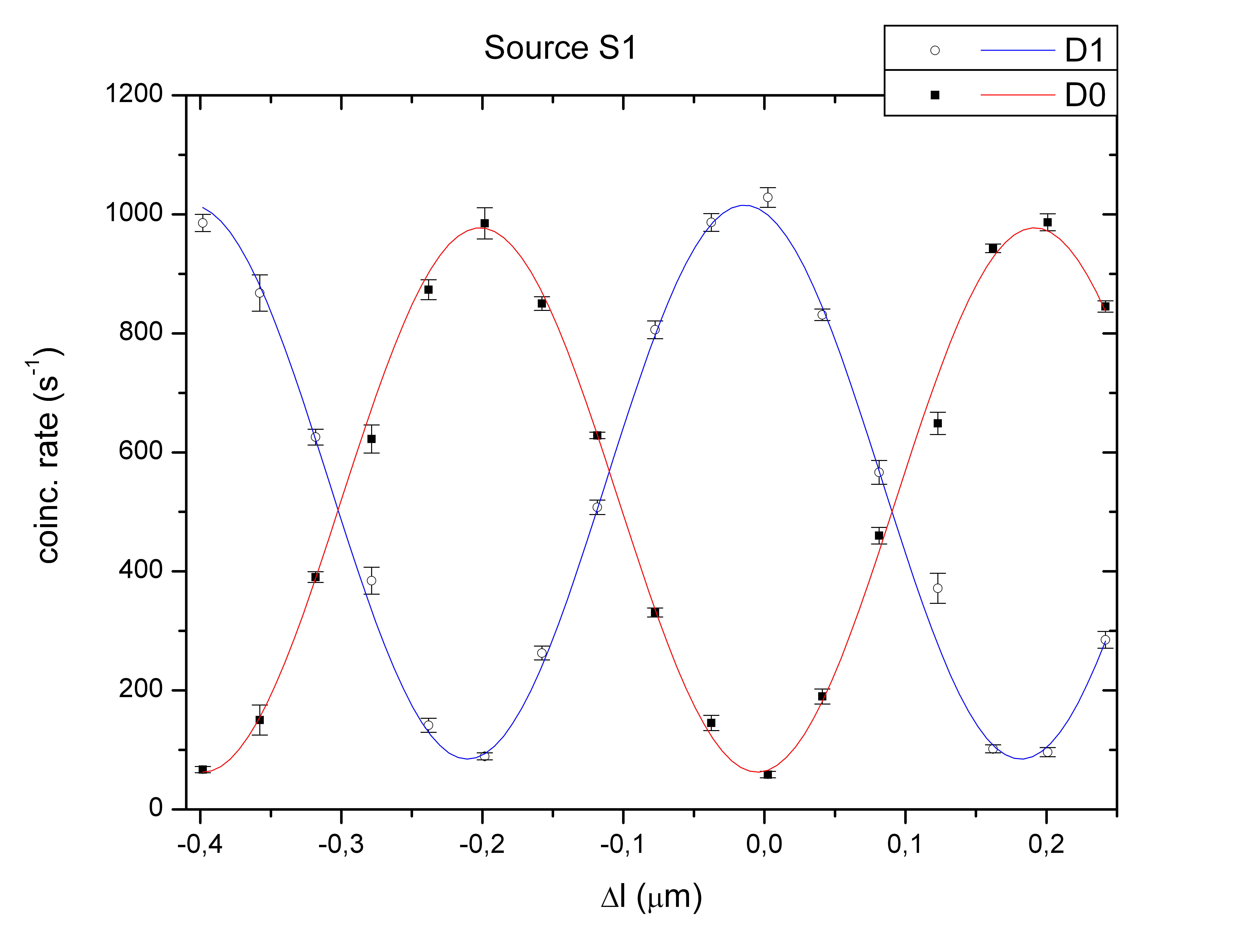}
   \end{center}
   \caption[example5]
   { \label{fig:example5} Number of detected events per second at detector
$D_0$ and $D_1$ 
as a function of the path
length difference $\Delta l$ 
between the two arms of the
interferometer for source $S_0$ (top picture) and $S_1$ (bottom). 
As expected, the phase
shifts between $D_0$ and $D_1$ sine fits of the coincidence counts are consistent with $\pi$.}
   \end{figure}

\begin{figure}
   \centering
    \includegraphics[width=9cm]{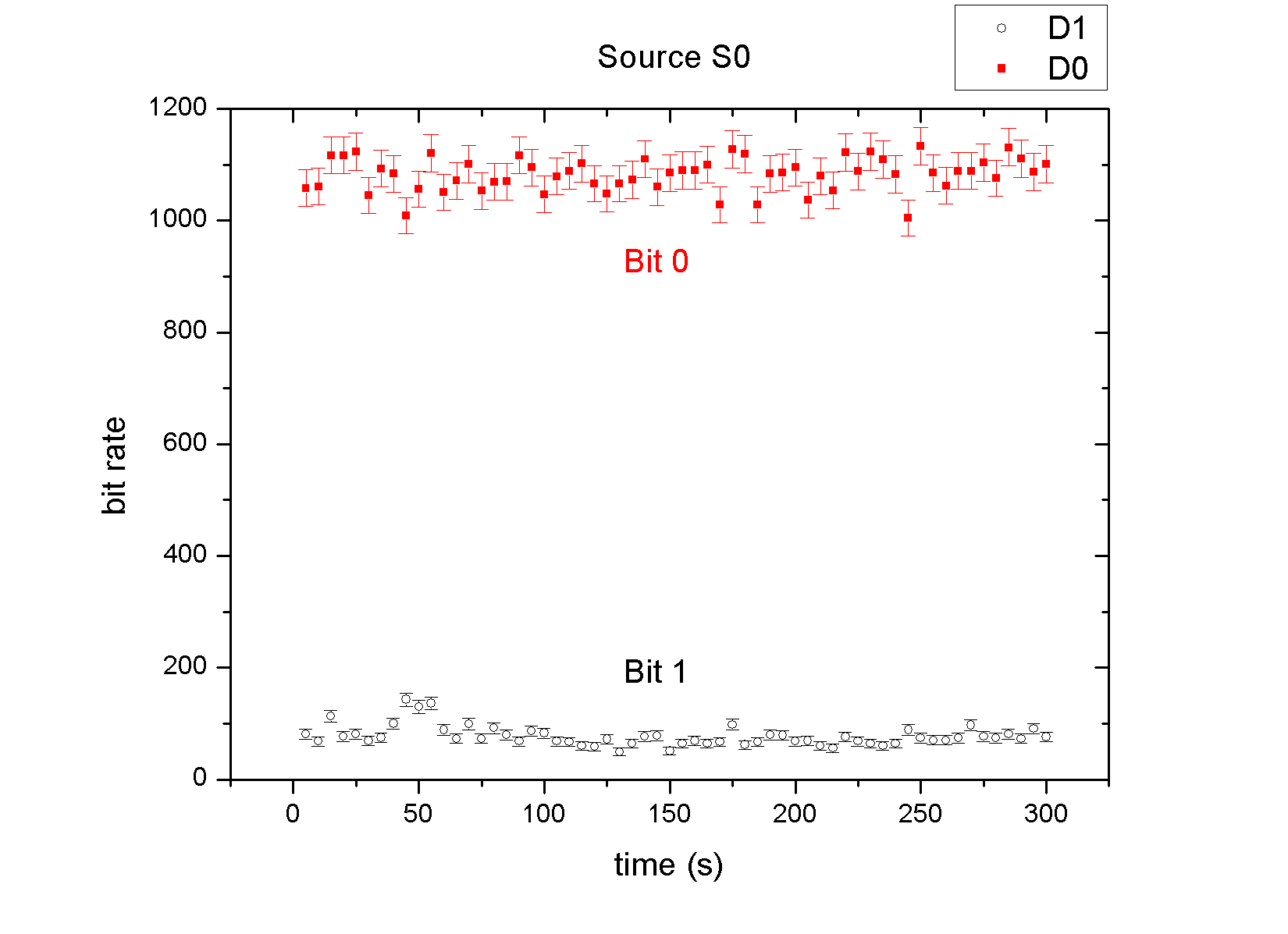}\\  \includegraphics[width=9cm]{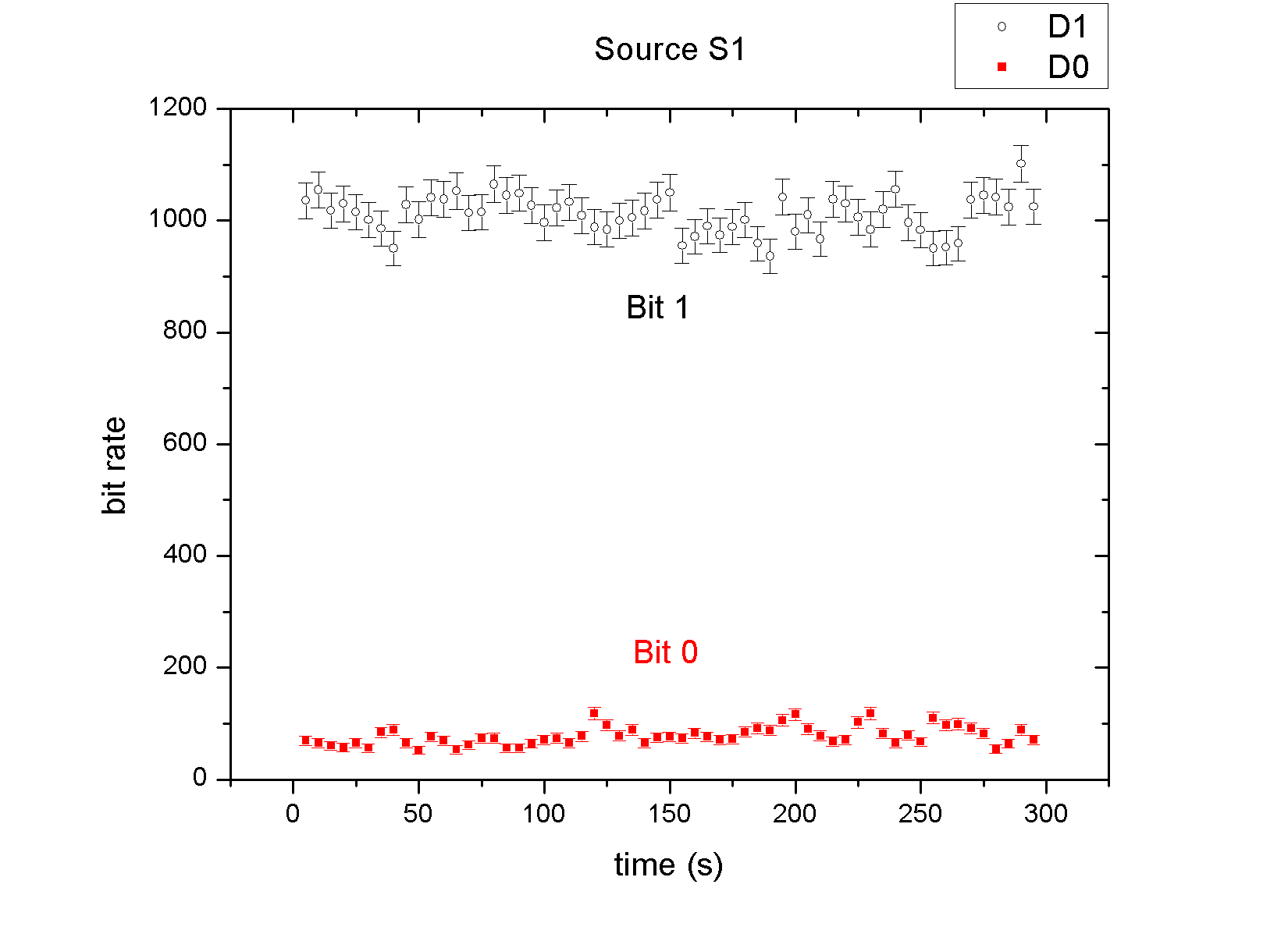}
   \caption[example6]
   { \label{fig:example6} Detection events at both detectors $D_1$ and $D_0$. Top: source $S_0$ is active,
   corresponding to the transmission of a string of bit $0$. Bottom: source $S_1$ is active, corresponding
   to the transmission of a string of bit $1$. The evaluated Quantum Bit Error Rate ($QBER$) in the two
   cases are $QBER_{S1}=0.071 \pm 0.014$ and $QBER_{S0} = 0.070 \pm 0.016$ on a series of $60$ measurements
   5 seconds long, showing a remarkable phase stability of the interferometer.}
   \end{figure}
The quality of the transmission is quantified by the Quantum Bit Error Rate
($QBER=\frac{P_{Wrong}}{P_{Right}+P_{Wrong}}$, where $P_{Right}$ ($P_{Wrong}$) is the probability for Bob to receive a bit value which is equal to (different from) the one sent by Alice), measured to be $7\%$, according to Table I.




\begin{table}[htbp]
\centering
   \begin{tabular}{|c||c|c|c|}
   \hline
    & $V_{D0}$ & $V_{D1}$ & $QBER$\\
   \hline
   \hline
   $S0$ & $(89 \pm 1)\%$ & $(82 \pm 1)\%$ & $(7.0 \pm 1.6)\%$\\
   \hline
   $S1$ & $(88 \pm 1)\%$ & $(85 \pm 1)\%$ & $(7.1 \pm 1.4)\%$\\
   \hline
   \end{tabular}
\caption{Main results obtained in the implementation of the QKD protocol proposed in Ref. \cite{g}.
$V_{D0}$, $V_{D1}$ are the visibilities of the interference fringes observed at the two outputs of the interferometer by scanning the path length difference, $QBER$ is the estimated quantum bit error rate
for the transmission
. }
   \end{table}

In conclusion, we have realized the first proof-of-principle experimental implementation of QKD
based on orthogonal states (GV protocol) 
\cite{g}. Our results
demonstrate the possibility of achieving a secure QKD transmission with orthogonal state and
therefore provides a significant hint to the discussion on the minimal quantum resources necessary
for the implementation of quantum tasks overcoming classical limits.

\acknowledgments{This work has been supported by PRIN 2007FYETBY (CCQOTS) and NATO (CBP.NR.NRCL 983251).
We thank L. Vaidman for having pointed out to our attention his
theoretical work.}


\end{document}